\documentclass[%
 reprint,
 amsmath,amssymb,
 aps,
]{revtex4-2}

\usepackage{graphicx}
\usepackage{dcolumn}
\usepackage{bm}
\usepackage{subcaption}

\begin{document}

\title{Catastrophe theoretic approach to the Higgs Mechanism}

\author{Samyak Jain}
\affiliation{Department of Physics, Indian Institute of Technology Bombay, 
Powai, Mumbai 400 076, India}%
\email{Corresponding author: samyakjain02@gmail.com}
\author{A. Bhagwat}%
\email{Contributing author: ameeya@cbs.ac.in}
\affiliation{School of Physical Sciences, UM-DAE Centre for Excellence in Basic Sciences, University of Mumbai, Kalina Campus, Mumbai 400 098, India}

\date{\today}

\begin{abstract}
A geometric perspective of the Higgs Mechanism is presented. Using 
Thom's Catastrophe Theory, we study the emergence of the Higgs Mechanism as a discontinuous feature in a general family of Lagrangians obtained by varying its parameters. We show that the Lagrangian that exhibits the Higgs Mechanism arises as a first-order phase transition in this general family. We find that the Higgs Mechanism (as well as Spontaneous Symmetry Breaking) need not occur for a different choice of parameters of the Lagrangian, and further analysis of these unconventional parameter choices may yield interesting implications for beyond standard model physics.
\end{abstract}

\maketitle


\section{Introduction} \label{introduction}
Catastrophe Theory is a geometrical framework developed to study sudden and discontinuous changes in dynamical systems under smooth perturbations. Thom showed (see \cite{thom}) that any smooth function of $n$ variables and $r$ parameters $(r \le 5)$ can be mapped to one and only one of 11 known families of functions (catastrophes). These catastrophes have unique geometries, and have already been studied in detail for sudden discontinuous changes under smooth perturbations to their parameters, thus allowing us to study \textit{any} $r$ parameter smooth function by finding mappings that take us to one of these known catastrophes (see \cite{poston} and \cite{arnold} for details). Catastrophe theory deals with systems that have these $r$ parameter functions as their potentials. We emphasize that these catastrophes are unique and cannot related to each other by smooth variable transformations.
 
Proposals to study quantum systems using Catastrophe theory have been suggested (see \cite{IBM}). Motivated by these, we hunt for discontinuous features typical of catastrophes in the Lagrangian demonstrating the Higgs Mechanism. In Section \ref{Catastrophe Theory}, we give a brief introduction to the cusp catastrophe (following the terminology of \cite{IBM}), which is one of the simplest but most commonly arising catastrophes. In Section \ref{higgs mech}, following \cite{griffiths}, we briefly outline how the Higgs Mechanism arises from a combination of Spontaneous Symmetry Breaking and local gauge invariance. In Section \ref{equiv cusp}, we show how the Lagrangian (its potential) used in Section \ref{higgs mech} is related to the Cusp Catastrophe, which describes a more general family of potentials. We show that attenuating the parameters of this general family leads to a first-order phase transition, as defined in \cite{IBM}. We also examine the the occurrence of the Higgs Mechanism and Spontaneous Symmetry Breaking in the larger parameter space spanned by the family of Lagrangians, and find that neither may occur for a general member of this family. \\

We must note that unlike the usual dynamical systems that Catastrophe theory is typically applied to, the parameters of the Higgs Lagrangian are fixed in nature, and cannot be adjusted and observed for changes. However, if the actual parameter values are even minutely away from the conventionally accepted parameter values, the results of our analysis become relevant for extending the standard model to these unconventional parameter choices . We motivate future investigation of this larger parameter space for insights into beyond standard model physics.

\section{The Cusp Catastrophe}
\label{Catastrophe Theory}
\begin{figure}[h]
\includegraphics[width=0.9\linewidth]{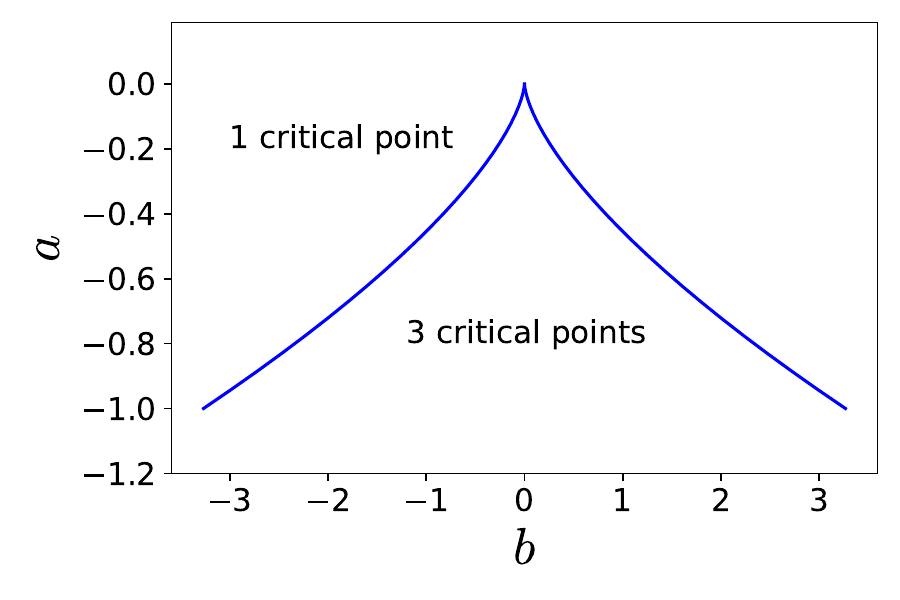}
 \caption{Variation in the number of critical points of the cusp potential in the $b-a$ plane. }
 \label{cusp fig}
\end{figure}

\begin{figure*}
\includegraphics[width=.9\linewidth]{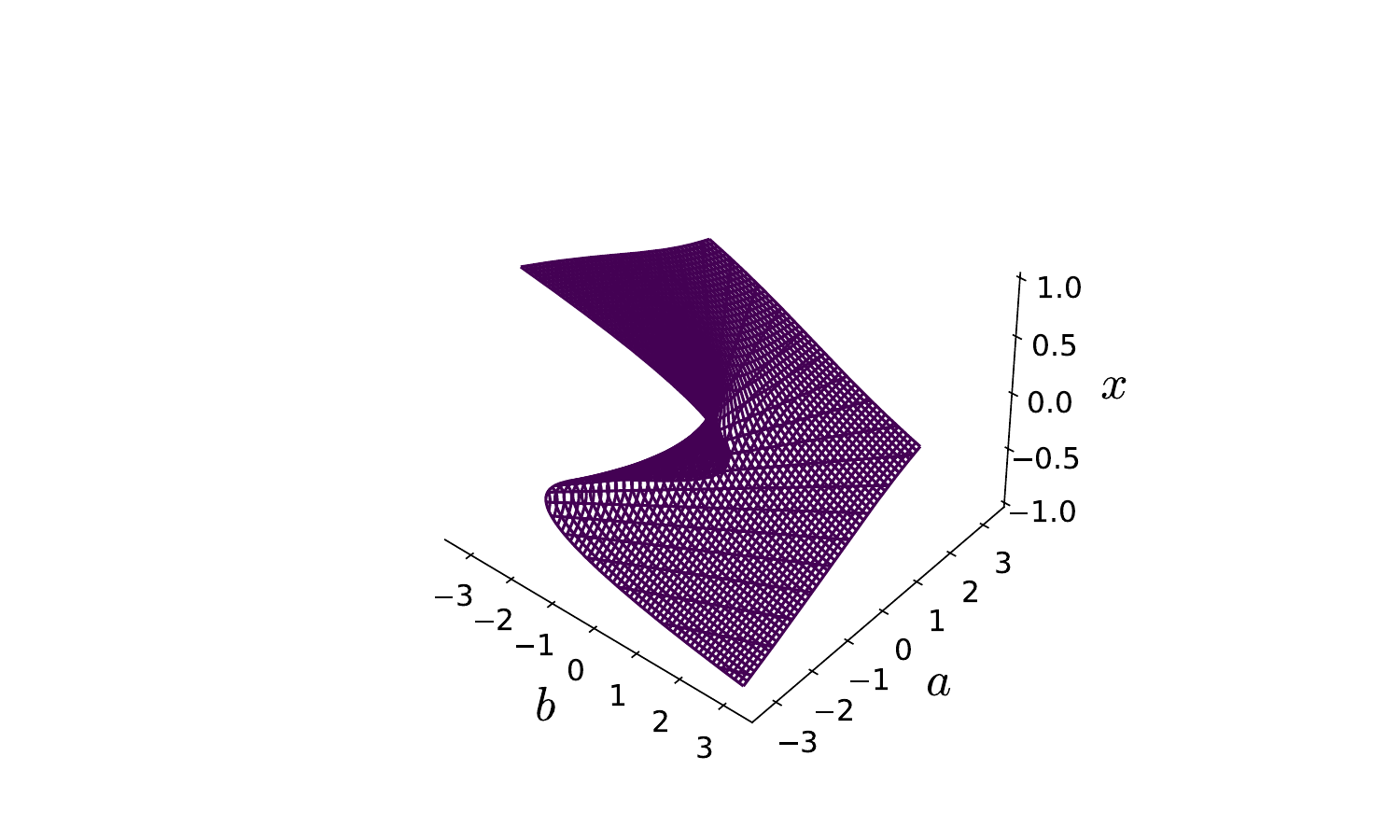}
\caption{\label{manifold} The catastrophe manifold for the Cusp Catastrophe. We can see how the critical points (number and values) vary across the $b-a$ plane.}
\end{figure*}

As observed earlier, the cusp catastrophe is an extremely common catastrophe in physical systems. The cusp potential reads as:
\begin{eqnarray*}
    V_\text{cusp}(a,b;x) = x^4 + ax^2 +bx, \quad x \in \mathbb{R} \label{cusp}
\end{eqnarray*}
with $a,b$ as parameters.
The critical points are defined as points where the first derivative of this potential vanishes, and satisfy
\begin{eqnarray}
    4x_c^3 + 2ax_c + b = 0 \label{crit}
\end{eqnarray}
This can be solved to see how the number of critical points $x_c$ varies with $a, b$. We obtain a single critical point (a local minimum) for all choices of parameters $a, b$ except for the region
\begin{eqnarray}
    a < 0, \quad |b| \le \frac{4}{3\sqrt{6}}\sqrt{-a^3}
\end{eqnarray}
where we have a local maxima sandwiched between 2 local minima. Fig.\ref{cusp fig} shows how this looks in the $b-a$ plane. In Fig.(\ref{manifold}) we show a clearer geometric picture by plotting Eq.(\ref{crit}) in the  $b-a-x$ space. This structure defines the so-called \textit{catastrophe manifold} (see \cite{poston}).

\begin{figure*}
\begin{subfigure}{0.3\textwidth}
\includegraphics[scale = 0.5]{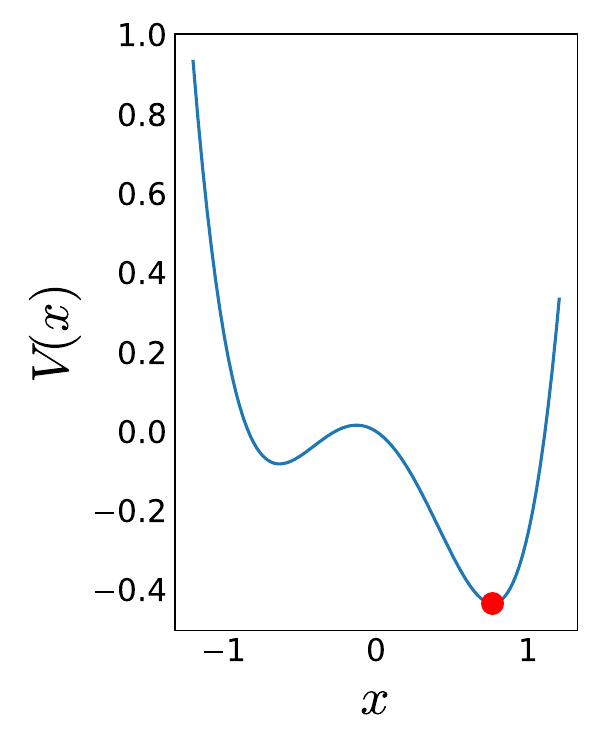}
  \subcaption{$a = -1, b = -\frac{1}{4}$}
\end{subfigure} 
\begin{subfigure}{0.3\textwidth}
  \includegraphics[scale = 0.5]{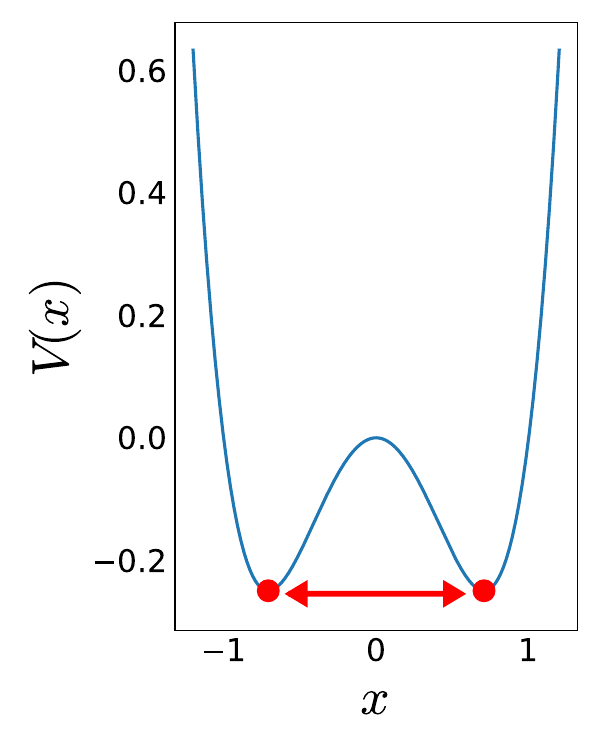}
  \subcaption{$a = -1, b = 0$}
  \label{phase transition_b=0}
\end{subfigure}
\begin{subfigure}{0.3\textwidth}
  \includegraphics[scale = 0.5]{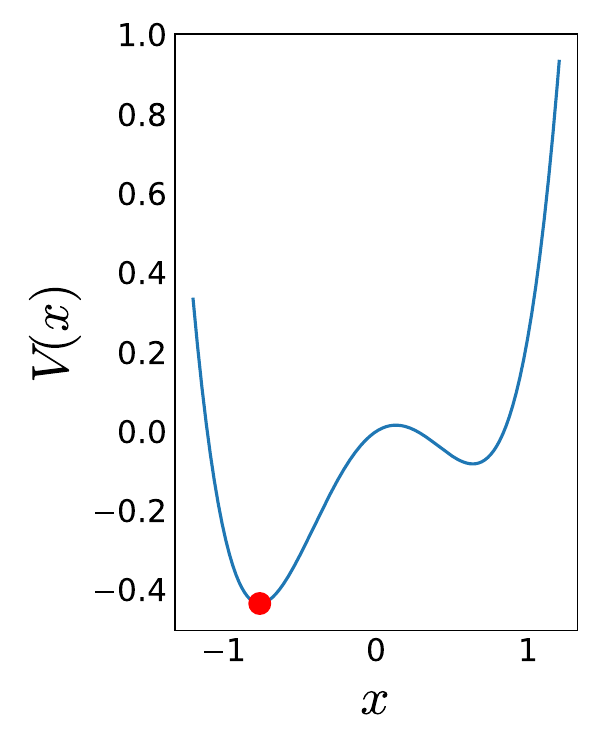}
  \subcaption{$a = -1, b = \frac{1}{4}$}
\end{subfigure}
\caption{\label{phase transition} The first-order phase transition in the cusp potential is shown. We fix $a = -1$, vary $b$ from $\frac{-1}{4}$ to $\frac{1}{4}$ and plot the cusp potential. The ground states for the potentials are indicated by red dots.
\newline
\newline 
a) For $b = -\frac{1}{4}$, we have two negative critical points, and the ground state is at the positive critical point (a minimum). b) At $b = 0$, the minima become degenerate, and the ground state can be at either minimum. Till now, the ground state energy has been decreasing with increasing $b$. c) As soon as $b > 0$, the ground state shifts abruptly to the negative minimum. The ground state energy now suddenly starts decreasing with increasing $b$. Since the ground state was never at $x = 0$, the rate of change of energy given by Eq.(\ref{deriv}) changes from a strictly positive value to a strictly negative value. Thus the rate of change of energy has shifted discontinuously, completing the first-order phase transition.}
\end{figure*}

As in \cite{IBM}, we assume that for any fixed choice of parameters, the energy of the system described by this potential dwells at the lowest minima of the potential. Consider $a < 0$ and $b$ varying from negative to positive values. We see that at $b = 0$, the minima become degenerate, and the lowest potential minimum shifts abruptly (see Fig. \ref{phase transition}). We can show that while the ground state energy $E_g$ is continuous (because the shift happens when the minima are degenerate), the rate of change of the ground state energy $\frac{dE_g}{db}$ is discontinuous:
\begin{eqnarray}
    \frac{dE_g}{db} = \frac{dE(x_c, a, b)}{db} = \frac{\partial{E(x_c, a, b)}}{\partial b} \nonumber\\
    + \frac{dE(x_c, a, b)}{dx_c} \frac{dx_c}{db}
\end{eqnarray}
Note that $x_c$ here is the particular critical point at which the ground state dwells. Now, the second term vanishes because $\frac{dE(x_c, a, b)}{dx_c} = 0$ by the definition of $x_c$. Thus
\begin{eqnarray}
    \frac{dE_g}{db} = x_c \label{deriv}
\end{eqnarray}
Since $x_c$ is discontinuous at $b = 0$, the first derivative of the ground-state energy is discontinuous as well. This is called a first-order phase transition (\cite{IBM}). \\
As shown in \cite{IBM}, we obtain a second-order phase transition (the second derivative being discontinuous) at $a = b = 0$. We do not describe this in detail because, as we shall see in Section \ref{equiv cusp}, we are concerned with only the first-order phase transition in this cusp potential.

\section{The Higgs Mechanism \label{higgs mech}}
We begin with the Mexican Hat Lagrangian that demonstrates spontaneous symmetry breaking \cite{griffiths}
\begin{eqnarray}
    \mathcal{L} = \frac{1}{2}(\partial_\mu \phi_1^*)(\partial^\mu \phi_1) 
    + \frac{1}{2}(\partial_\mu \phi_2^*)(\partial^\mu \phi_2)  \nonumber\\ + \frac{\mu}{2}(\phi_1^2 + \phi_2^2) - \frac{\lambda}{4}(\phi_1^2 + \phi_2^2)^2 \label{ssb lag}
\end{eqnarray}
We define
\begin{equation}
    \phi = \phi_1 + i\phi_2, \quad \phi^* = \phi_1 - i\phi_2
\end{equation}
The Lagrangian thus becomes
\begin{equation}
    \mathcal{L} = \frac{1}{2}(\partial_\mu \phi)^*(\partial^\mu \phi) + \frac{\mu}{2}(\phi^* \phi) - \frac{\lambda}{4}(\phi^*\phi)^2 \label{special lag}
\end{equation}

\begin{figure}[h!]
\includegraphics[scale=0.65]{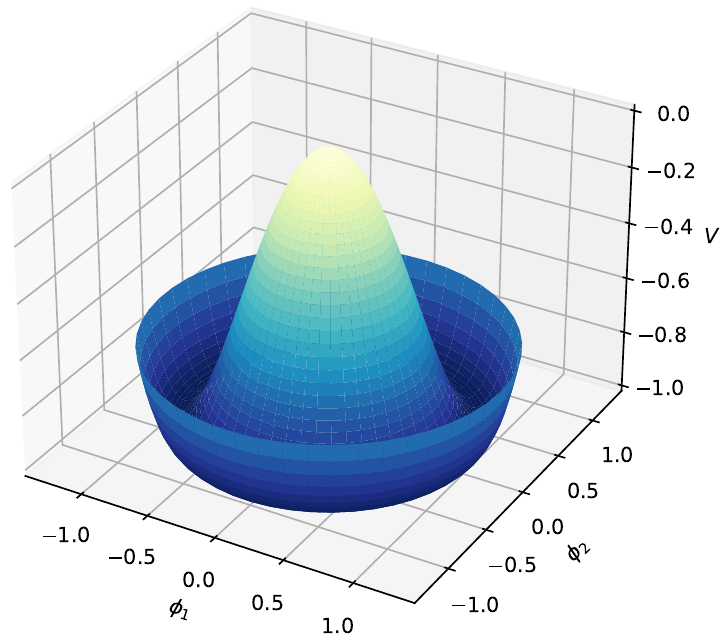}
 \caption[The Higgs potential is plotted in the $\phi_1-\phi_2$ plane]{The Higgs potential defined by Eq.(\ref{potential}) is plotted in the $\phi_1-\phi_2$ plane for $\lambda = \mu = 4$. Note the resemblance to Fig.\ref{phase transition_b=0}.}
 \label{mexican hat}
\end{figure}

Enforcing local gauge invariance, we introduce a field $A^\mu$ that transforms like 
\begin{equation}
    A^\mu \rightarrow A^\mu + \partial^\mu \lambda
\end{equation}
under a transformation \begin{equation}
    \phi \rightarrow e^{i\theta(x)}\phi \label{lgv}
\end{equation}
We further complete the Proca Lagrangian by introducing the term $\frac{-1}{16\pi}F^{\mu \nu}F_{\mu \nu}$. Finally, we replace partial derivatives with covariant derivatives to obtain the locally gauge-invariant Lagrangian:
\begin{eqnarray}
        \mathcal{L} = \frac{1}{2} \left( \left[ (\partial_\mu - iqA_\mu \right]\phi^*\right)\left( \left[ \partial^\mu + iqA^\mu)\right]\phi \right) \nonumber\\
    + \frac{1}{2}\mu^2(\phi^*\phi) - \frac{1}{4}\lambda^2(\phi^*\phi)^2 - \frac{1}{16\pi}F^{\mu \nu}F_{\mu \nu}
    \label{Higgs Lag}
    \end{eqnarray}
    The Higgs Mechanism is seen by expanding this Lagrangian density about the minima (ground state) of the potential of the initial Lagrangian 
    :
    \begin{eqnarray}
                V = \frac{\lambda}{4}(\phi^*\phi)^2 - \frac{\mu}{2}(\phi^*\phi) \nonumber\\
        =\frac{\lambda}{4}(\phi_1^2 + \phi_2^2)^2 - \frac{\mu}{2}(\phi_1^2 + \phi_2^2) 
        \label{potential}
    \end{eqnarray}
    The ground state is obtained at
    \begin{eqnarray}
        \phi^* \phi = \frac{\mu^2}{\lambda^2}
    \end{eqnarray}
    instead of $\phi_1 = \phi_2 = 0$; this can be seen in Fig.\ref{mexican hat}, where we plot the Higgs potential defined by Eq.(\ref{potential}).
    
    We conveniently define
    \begin{eqnarray}
        \phi_1 = \eta + \frac{\mu}{\lambda}, \phi_2 = \xi 
    \end{eqnarray}
    As shown in \cite{griffiths}, upon rewriting the Lagrangian in Eq. (\ref{Higgs Lag}) in terms of these new fields we obtain a new particle defined by the field $\eta$ (the Higgs) and other interaction terms. A Goldstone Boson ($\xi$) is obtained as well, along with a bilinear term
    \begin{eqnarray}
        -2i\left( \frac{\mu}{\lambda} \frac{q}{\hbar c} \right) \left( \partial_\mu \xi\right)(A^\mu)
        \label{int term}
    \end{eqnarray}
    This term cannot be interpreted as an interaction term, because it implies that the particles defined by $\xi$ and $A^\mu$ cannot exist independently. To get rid of the Goldstone Boson and this bilinear term, \cite{griffiths} exploits the local gauge invariance of the Lagrangian defined by Eq.(\ref{Higgs Lag}), and chooses 
    \begin{eqnarray}
        \theta = -\tan ^{-1} \left( \frac{\phi_2}{\phi_1} \right)
    \end{eqnarray}
 In this gauge, $\phi_2$ trivially vanishes, thus eliminating both the Goldstone Boson and the bilinear term. This completes the Higgs Mechanism. \\
 We emphasize that at no step was the initial Lagrangian given by Eq.(\ref{Higgs Lag}) altered, it was only redefined by simple re-parameterization of the initial fields $\phi_1$ and $\phi_2$, and a suitable $\theta$ was chosen at the end. At any point, we can revert to the initial choice of fields and arrive at the initial Lagrangian. The key features of the Mechanism are the emergence of the Higgs ($\eta$), elimination of the Goldstone Boson ($\xi$), and the non-physical interaction term (Eq.(\ref{int term})) when the Lagrangian is expanded about its potential's minimum and a suitable gauge is chosen.

\section{Mapping to the Cusp Potential}
\label{equiv cusp}
In this section, we study the rather unique choice of Lagrangian defined by Eq.(\ref{special lag}). The Higgs Mechanism as detailed above is a very specific phenomenon, and need not arise in a more general Lagrangian. We thus now apply a catastrophe theoretic approach to the potential defined by Eq.(\ref{potential}).

We see that the potential only depends on $\phi^*\phi$, and effectively is a one-variable function. We define $r = \sqrt{\phi^* \phi}$ and obtain:
\begin{equation}
    V = \frac{\lambda^2}{4}r^4 - \frac{\mu}{2}r^2
\end{equation}

This potential can be mapped to the cusp potential defined by Eq.(\ref{cusp})
via  
\begin{eqnarray}
    x = \sqrt{\frac{\lambda}{2}} \, r, \quad a = -\frac{\mu}{\lambda}, \quad b = 0
\end{eqnarray}
 We note that this automatically assigns a unique geometry (that of the Cusp Catastrophe) to this family. The family cannot be mapped to any other catastrophe. As emphasized in Section \ref{introduction}, each Catastrophe has a unique geometry, and thus a family of functions can never be mapped to more than one catastrophe. All the re-parameterizations of the initial Lagrangian in Section \ref{higgs mech} can be mapped to (only) the Cusp Catastrophe, simply because the initial Lagrangian maps to the Cusp Catastrophe.\\
We immediately see that we indeed have something special happening. We have already shown in Section \ref{Catastrophe Theory} that we have a first order-phase transition at $b = 0$ for a fixed $a<0$, which is clearly true for our potential. This means that the potential that gave rise to the Higgs Mechanism in Section \ref{higgs mech} can be generalized by adding a term $br$ with $b \ne 0$. Thus the Higgs Mechanism arises with a first-order phase transition in the family of Lagrangians with potentials given by
\begin{eqnarray}
    V_f = \frac{\lambda}{4} r^4 - \frac{\mu}{2}r^2 + br, b \in \mathbb{R}
    \label{new_pot}
\end{eqnarray}
It is easy to verify that the Higgs Mechanism cannot show up with the new term 
\begin{eqnarray}
    br = b\sqrt{\phi^* \phi} = \sqrt{\left( \frac{\mu}{\lambda} + \eta \right) ^2 + \xi^2}
\end{eqnarray} 
There is no way to interpret this term as an interaction. The fields cannot be promoted to operators under canonical quantization. We cannot exploit the local gauge invariance of $\mathcal{L}$ to get rid of this term, $\sqrt{\phi^*\phi}$ is trivially invariant under the transformation defined by Eq.(\ref{lgv}), and vanishes only for $\phi_1 = \phi_2 = 0$. Thus the Higgs mechanism does not show up in this general Lagrangian family, it arises suddenly only when $b = 0$. \\

\begin{figure*}
\begin{subfigure}{0.32\textwidth}
\centering
\includegraphics[scale = 0.52]{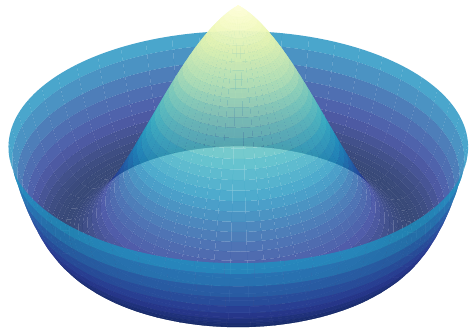}
  \vspace{0.5cm}

  \subcaption{$b=-1$}
\end{subfigure} 
\begin{subfigure}{0.32\textwidth}
  \includegraphics[scale = 0.52]{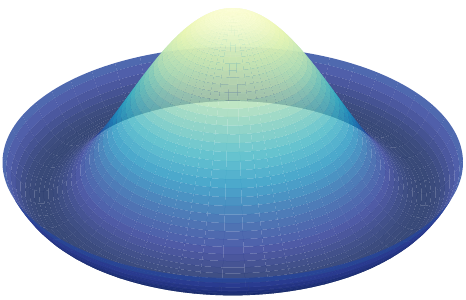}
  \vspace{0.5cm}
  \subcaption{$b = 0$}
\end{subfigure}
\begin{subfigure}{0.32\textwidth}
  \includegraphics[scale = 0.52]{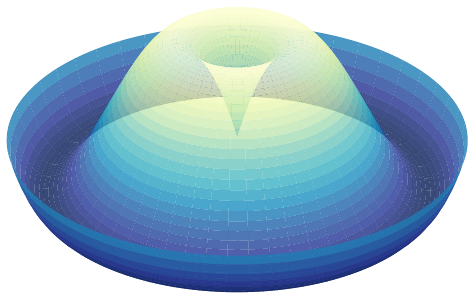}
    \vspace{0.5cm}

  \subcaption{$b = 1$}
\end{subfigure}
\caption[The general shape of new Higgs potential is plotted for different values of $b$.]{\label{ssb broken}The new Higgs potential (Eq.(\ref{new_pot}) is plotted for different values of $b$, with fixed $\lambda = \mu = 4$. The axes are hidden due to a different scaling of each figure to highlight relevant features. Note the cone-like structure at the origin for the non-zero values of $b$, and how the origin acts as a minimum for $b>0$ despite the first derivative with respect to $r$ being non-zero.}
\end{figure*}

We can also investigate the occurrence of Spontaneous Symmetry Breaking in this family of Lagrangians by plotting the potential as we move around in the parameter space; see Fig.\ref{ssb broken}

We see that upon introducing a non-zero $b$, the $(0,0)$ point becomes the vertex of a cone-like structure; this is an artifact of the constraint $r= \sqrt{\phi^* \phi} \ge 0$. Particularly, for $b>0$, the cone is inverted, and the vertex behaves as a minimum. This implies that despite the first derivative of the potential with respect to $r$ being non-zero, the origin \textit{can} act as a ground state. In fact, for large enough positive values of $b$, the origin is the \textit{lowest} minimum. If the system's ground state resides here, then Spontaneous Symmetry Breaking itself is \textit{broken}. This implies from Eq.(\ref{ssb lag}) that the fields have an \textit{imaginary} mass, which is unphysical. \\

There are two possible interpretations of this observation. One is that this simply makes our argument stronger, that we \textit{must} have $b = 0$ for the Higgs Mechanism to occur. The second, more intriguing explanation is that there is some new physics at play for the $b>0$ case. For the $b<0$ case, the origin is still a `vertex' but will act as a local maximum instead, and cannot be the ground state. Spontaneous Symmetry Breaking is thus observed in this case, but the Higgs Mechanism is not due to the abnormal interaction term seen earlier. \\



\vspace{0.3cm}

\section{Summary and Conclusion}
We present a catastrophe theoretic interpretation of the Higgs Mechanism. We showed that the Lagrangian demonstrating the Higgs Mechanism (Eq.(\ref{special lag})) is a special member of a general family of Lagrangians. The potential of this family of Lagrangians can be easily mapped to the cusp catastrophe, thus assigning a unique geometry to the family.
We find that the family demonstrates a first-order phase transition (as defined in Section \ref{Catastrophe Theory}) when its parameters are varied continuously so as to arrive at the usual Lagrangian exhibiting the Higgs Mechanism. We also find that a general member of this family of Lagrangians cannot show the Higgs Mechanism, and that Spontaneous Symmetry Breaking is not observed for a large enough positive $b$ . \\

We note here again that the parameter values of the Higgs Lagrangian are fixed in nature $(b = 0)$. However, one may instead have small non-zero values of $b$ as well. The full implications of our analysis in the context of beyond standard model physics are yet to be explored.


\vspace{0.5cm}

\section{Acknowledgements}
We would like to thank Urjit A. Yajnik for his valuable insight into our work. We thank Ajay Patvardhan for his suggestions.



\providecommand{\noopsort}[1]{}\providecommand{\singleletter}[1]{#1}%

\end{document}